\documentclass[aps,prb,twocolumn]{revtex4}
\usepackage{times,amsfonts,amsmath,amsthm,amssymb,graphicx,mdframed,siunitx,lettrine,caption,fancyhdr,pifont,mathtools,ragged2e}
\usepackage[T1]{fontenc}
\usepackage[export]{adjustbox}
\newcommand{\newtextbar}[1][2]{\scalebox{#1}[1]{\textbar}}
\DeclareCaptionLabelSeparator{bar}{ \newtextbar \;}
\newcommand{\mli}[1]{\mathit{#1}}
\definecolor{oranj}{RGB}{236,122,0}
\definecolor{laciv}{RGB}{0,32,96}
\begin{document}
\title{\includegraphics[width=1\textwidth,left]{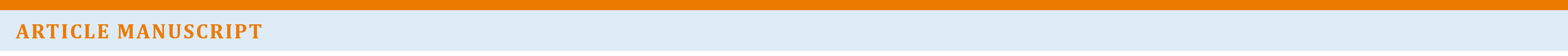}~ \flushleft \Large {\fontfamily{phv}\selectfont
Quantum shape effects and novel thermodynamic behaviors at nanoscale\\}
\flushleft{\normalsize Alhun Aydin \& Altug Sisman}\vspace{-4pt}}
\email{altug.sisman@physics.uu.se}
\begin{abstract}
\flushleft{\textit{Department of Physics and Astronomy, Uppsala University, 75120, Uppsala, Sweden \\
Nano Energy Research Group, Energy Institute, Istanbul Technical University, 34469, Istanbul, Turkey}}\\
\noindent{\color{oranj}\rule{\textwidth}{0.9pt}}\vspace*{-\baselineskip}\vspace*{1pt}
\noindent{\color{oranj}\rule{\textwidth}{0.2pt}}\vspace{3pt}
\textbf{{\fontfamily{lmss}\selectfont Thermodynamic properties of confined systems depend on sizes of the confinement domain due to quantum nature of particles. Here we show that shape also enters as a control parameter on thermodynamic state functions. By considering specially designed confinement domains, we separate the influences of quantum size and shape effects from each other and demonstrate how shape effects alone modify Helmholtz free energy, entropy and internal energy of a confined system. We propose an overlapped quantum boundary layer method to analytically predict quantum shape effects without even solving Schr\"odinger equation or invoking any other mathematical tools. Thereby we reduce a thermodynamic problem into a simple geometric one and reveal the profound link between geometry and thermodynamics. We report also a torque due to quantum shape effects. Furthermore, we introduce isoformal, shape preserving, process which opens the possibility of a new generation of thermodynamic cycles operating at nanoscale with unique features.}}\vspace{-2pt} 
\noindent{\color{oranj}\rule{\textwidth}{0.25pt}}\vspace*{-\baselineskip}\vspace*{1.7pt}
\noindent{\color{oranj}\rule{\textwidth}{1pt}}
\end{abstract}
\maketitle
\small


\lettrine[findent=2pt]{\textbf{A}}{}dvances in quantum and nanotechnologies in recent years are so excessive that bringing quantum world into daily life starts to loom on the horizon. Exploration of quantum effects is a fascinating quest more than ever, as it may help to design novel devices with features beyond the current cutting edge technologies \cite{qtechh,class,macror}. To this objective, understanding and manipulating thermodynamic behaviors of nanoscale systems are crucial for the development of new devices at nanoscale. In recent years, numerous studies from both theoretical and experimental sides are conducted in the field of quantum and nano thermodynamics and already, various types of quantum heat engines, motors and nanoscale thermodynamic devices have been proposed \cite{naturerot1,naturerot2,PhysRevLett.102.230601,eplrot,anrevw,neqqhm,signat,PhysRevLett.111.060802,gyrator,naturerot3,flywheel,scirot,entkos}.

When quantum particles are confined in domains on the order of their de Broglie wavelength, quantum size effects (QSE) appear. At nanoscale, changing the size of a domain alters the thermodynamic properties of particles confined within \cite{baltes,pathria}. Lower dimensional geometric elements of a domain such as surface area ($A$), peripheral length ($P$) and even the number of vertices ($N_V$) enter as control parameters on thermodynamic state functions at nanoscale in addition to volume ($V$) \cite{baltes,pathria,discnat,ddos}. By controlling these size variables it is possible to manipulate and tailor the certain features of confined systems (e.g. semiconductors, metals, superconductors, topological insulators and ultracold gases) in a desired way \cite{rodun,rmpqse,qsenat,suprqsesc,daixie,nanscsh,qsewff,emequse,bcsbecqse,qsenjpg,qsescirp}. QSE give rise also to phenomena such as thermosize effects and specific heat oscillations which can be used in nanoscale energy conversion and storage technologies \cite{tse1,qoscnat,tse2,discnat}. Similar to QSE, difference in boundary conditions also lead to pressure forces of a quantum statistical origin due to the modification of energy spectrum \cite{flopb2,PhysRevE.83.041133,PhysRevA.82.062107}.

A pioneering work has been done by Hermann Weyl in 1911 \cite{weyl11}, about the asymptotic behavior of the eigenvalues of a Laplace operator for the domains with Dirichlet boundary conditions. Even though the study was about size/boundary effects on eigenvalue spectrum, it initiated the field of spectral geometry and the discussion of impact of shape on the physical properties of a system. The topic was popularized by Mark Kac with the question "Can one hear the shape of a drum?" \cite{kac66} as an inverse problem. Although the answer to this question is no\cite{ansno} in a mathematical sense, since there are few isospectral domains with different shapes. However, these type of domains are actually extremely rare and very specially arranged \cite{revcanone} and for most systems, shape of the boundaries indeed determines the eigenvalue spectrum\cite{revisit,shapeDNA1,sciemano}. Like size effects, influence of shape effects become also important when domain is so confined that particles reveal their quantum character. Although there are some studies in literature mentioning the phrase "quantum shape effects", none of them are related with thermodynamics and all of them in fact study the size and shape effects together without considering the shape effects alone, since they define the shape by aspect ratios or geometric structure (\textit{e.g.} rectangular, circular) of the domain \cite{infshapetr,shapecasimir,biss,shapemat,shapeeffnl,pchenprb,nantech,nanscsh,langmu,pccp1}. However, in those cases, at least one of the size variables ($V, A, P, N_V$) also changes and it's not possible to analyze shape effects solely. The pure shape dependence on thermodynamic behaviors of confined systems has never been examined before.

In the present article, we first determine a special type of confinement domain where the shape can be changed without altering geometric size elements so that shape effects are separated from size effects. Then, we show how quantum shape effects (QShE) arise and change Helmholtz free energy, entropy and internal energy of the confined system. Overlapped quantum boundary layer method, proposed here, provides a solid framework to predict and explain the underlying mechanisms of QShE. We reveal the profound link between thermodynamics and geometry by showing that minimization of free energy under isothermal process is actually equivalent to the maximization of effective volume instead of volume itself. We obtain analytical expressions predicting the shape dependence of thermodynamic state functions by considering overlapping quantum boundary layers. Some thermodynamic properties like entropy, behaves unexpectedly under shape variation processes. Occurrence of the simultaneous decrement in entropy and free energy is also explained by the detailed examination of system's behaviors under QShE. QShE on thermodynamics gives possibility to define novel thermodynamic processes in quantum realm. We show the possibility of exchanging heat and work without changing volume and all other size parameters during an isothermal process. We introduce a completely new type of power cycle based on QShE, consisting two isothermal and two isoformal (shape preserving) processes. In addition, a torque effect reveals as a consequence of the inhomogeneous pressure distributions of particles along the domain walls due to QShE. The torque or QShE in general may be seen as a new macroscopic manifestation of quantum phenomenon since the effects persist even if sizes of the domain in one or two directions are much larger than de Broglie wavelength of particles.

\text{ }\\
\noindent\textbf{\normalsize{Results}}\\
\noindent\textbf{Shape as a new thermodynamic control variable.}
As it may be considered a somewhat vague concept, it's necessary to define what do we mean by shape in the first place. Here we consider the shape as geometric information of an object invariant under Euclidean similarity transformations such as translation, rotation, reflection and uniform scaling \cite{shape1,shape2}.

An effect can be called shape effect only if it depends purely and simply on the characteristic shape of the system. Generally in literature, aspect ratios or geometric structures are considered as indicators of shape characteristics of a domain \cite{biss,shapemat,pchenprb,nantech,nanscsh,langmu,pccp1}. However, when aspect ratio of a domain changes, size variables also change along with shape. Similarly, a geometric structure change (\textit{e.g.}. from cube to sphere) cannot be done without changing geometric size variables of a domain, \textit{i.e.}. $V$, $A$, $P$ and $N_V$. In other words, in those type of processes size and shape effects are inherently linked and cannot be separated from each other.

To investigate the pure shape effects on thermodynamic properties of confined systems, we need a confinement domain whose all geometric size variables ($V$, $A$, $P$ and $N_V$) remain constant under a shape transformation. For this purpose, the confinement domain that we choose consists of two coaxial square rigid nanowires having different edge lengths placed inside one another. The outer wire is fixed whereas the inner one is free to rotate and the space between two wires is occupied by particles. Boundaries are taken to be impenetrable. Cutaway views of the confinement domain for two different rotational configurations of inner wire are shown Fig. 1. Here, shape transformation is characterized by the rotation angle $\theta$ (clockwise direction is taken to be positive) of the inner square wire. Different angular configurations correspond to different confinement shapes perceived by the confined particles, while the geometric size variables of the domain remain the same. Note that particles are confined inside a three-dimensional (3D) domain, but for convenience two-dimensional (2D) cutaway views of confinement domains are shown in all figures throughout the article.

\begin{figure}[t]
\centering
\includegraphics[width=0.35\textwidth]{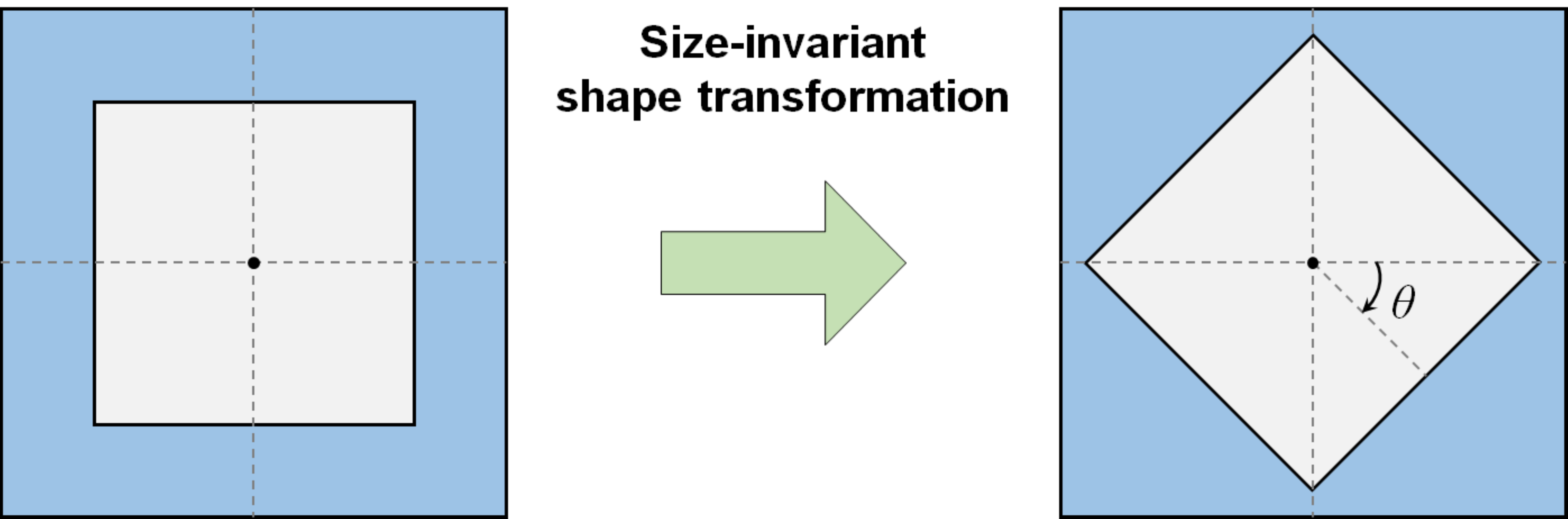}
\caption{\bfseries{Size-invariant shape transformation.} \mdseries Cutaway views of two nested square nanowires with fixed outer wire and rotatable inner wire. Particles are confined in a space between two wires denoted by blue region. The confinement domain undergoes to a size-invariant shape transformation by virtue of the rotation of inner square wire. Shape transformation is characterized by the clockwise variation of rotation angle $\theta$.}
\label{fig:pic1}
\end{figure}

The domain is chosen as strongly confined in transverse directions to reveal QShE, whereas nearly free in longitudinal direction. Thus, there is always enough number of particles in the confinement domain to use statistical methods even in low density conditions which allow us to use Maxwell-Boltzmann statistics. We quantify the strength of confinement in transverse and longitudinal directions by introducing respective confinement parameters as $\alpha_t=L_c/L_{*}=1$ and $\alpha_l=L_c/L_l=0.005$. Here, $L_c=h/\sqrt{8mk_BT}$ is the half of the most probable de Broglie wavelength of particles and $L_l$ is the size of the domain in longitudinal direction. $L_{*}$ is a characteristic length of the domain in transverse direction which is defined here as $L_{*}=2A_C/P_C$ where $A_C$ and $P_C$ are area and periphery of the domain's cross section. It corresponds to the harmonic mean size of the domain in transverse direction. Particle density and temperature are choosen as $n_{cl}=5\times 10^{18} \text{cm}^{-3}$ and $T=300\text{K}$ respectively in the calculations of thermodynamic state functions.

We first calculate the partition function for particles confined in the domain. Partition function contains a summation over discrete set of energy eigenvalues from the solution of the Schr\"odinger equation. We numerically solve the Schr\"odinger equation and obtain eigenvalues, since there is no analytical solution for our custom-shaped domain. Owing to the orthogonality of eigenstates, quadratic dispersion relation and product rule of exponents, it's possible to decompose partition function into the products of transverse and longitudinal parts. Therefore, instead of solving the Schr\"odinger equation for the 3D domain, we solve it for its 2D cross section and use the eigenvalues obtained from this numerical solution in the transverse part of partition function (see Methods section for the details of numerical calculations). For longitudinal part, we can easily obtain a precise analytical expression containing QSE corrections, by using the first two terms of Poisson summation formula (PSF) or equivalently by Weyl conjecture \cite{baltes,ddos}. By multiplying longitudinal and transverse parts obtained from analytical and numerical approaches respectively, the partition function reads
\begin{equation}
\zeta=\sum_\varepsilon{\exp\left(-\frac{\varepsilon}{k_BT}\right)}=\left(\frac{\sqrt{\pi}}{2\alpha_l}-\frac{1}{2}\right)\sum_k \exp\left(-\frac{\alpha_t^2}{\pi^2}\tilde{k}^2\right),
\end{equation}
where $\varepsilon$ denotes energy eigenvalues, $\tilde{k}=k L_{*}$ and $k$ is wavenumber. It should be noted that, the summations in Eq. (1) take possible degeneracies of eigenvalues into account also.

We now calculate Helmholtz free energy of the system by $F=-k_B T\ln Z$, where $Z=\zeta^N/N!$, for each degree of configuration from $0^{\circ}$ to $45^{\circ}$. Even though all conventional thermodynamic state variables such as density ($n$), temperature ($T$) and geometric size variables of the domain ($V, A, P, N_V$) are invariant under shape transformation, we obtain different Helmholtz free energy values for each different angular configuration (corresponding to different shapes), see solid black curve in Fig. 2a. Thermodynamic entropy ($S=-\partial F/\partial T$) and internal energy ($U=F+TS$) also depend on the rotation angle $\theta$ and change accordingly as they are shown by solid blue and solid red curves in Figs. 2d and 2g respectively. Free energy, entropy and internal energy decrease as the inner square wire rotates from starting position of $\theta=0^{\circ}$ to the $\theta=45^{\circ}$.

\begin{figure}[t]
\centering
\includegraphics[width=0.48\textwidth]{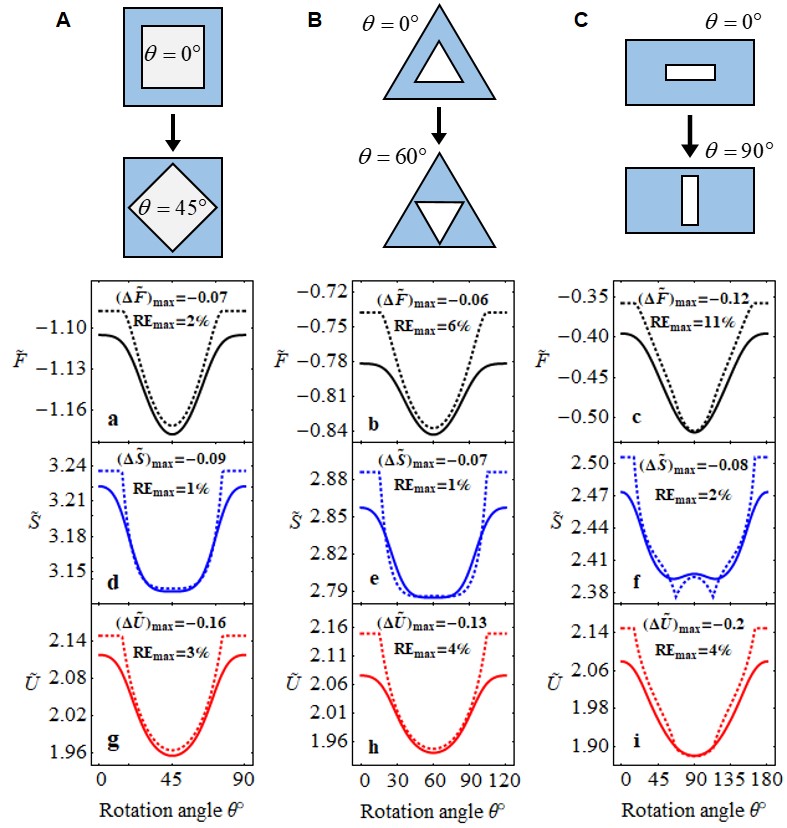}
\caption{\bfseries{Shape dependent thermodynamic state functions.} \mdseries Dimensionless Helmholtz free energy, $\tilde{F}=F/Nk_BT$, (\textbf{a}, \textbf{b}, \textbf{c}), entropy, $\tilde{S}=S/Nk_B$, (\textbf{d}, \textbf{e}, \textbf{f}) and internal energy, $\tilde{U}=U/Nk_BT$, (\textbf{g}, \textbf{h}, \textbf{i}) change with domain shape that is characterized by the rotation angle of inner objects, $\theta$. Columns \textbf{A}, \textbf{B}, \textbf{C} represent different confinement domain shapes and the thermodynamic quantities of particles confined in these domains. Solid and dotted curves represent numerical and analytical results respectively. For a thermodynamic quantity $\tilde{X}$, $(\Delta\tilde{X})_{\mli{max}}$ values inside the figures show the maximum change in each thermodynamic function based on solid curves. $RE_{\mli{max}}$ denotes the maximum relative errors of analytical predictions at each case.}
\label{fig:pic2}
\end{figure}

The variation of thermodynamic state functions with respect to rotation angle shows that shape of the system enters as a new control parameter on thermodynamic state functions which is characterized by $\theta$ variable. For a given number of particles ($N$) in a system, thermodynamic state space becomes six-dimensional, while it is two-dimensional in a simple system where QSE and QShE are negligible,
\begin{equation*}
X_N\left(T,V\right)\xrightarrow{\mli{QSE}}X_N\left(\ldots,A,P,N_V\right)\xrightarrow{\mli{QShE}}X_N\left(\ldots,\theta\right),
\end{equation*}
where $X_N$ is any thermodynamic state function of a system with particle number $N$.

The column denoted by A in Fig. 2 shows the shape dependence of thermodynamic state functions for our original confinement domain presented in Fig. 1. Besides the nested square confinement domain that is initially chosen, nested triangular (Fig. 2B) and nested rectangular (Fig. 2C) domains are considered as well. For the sake of comparability, the same particle density, transverse and longitudinal confinement parameters are chosen also for nested triangle and rectangle domains. Similar behaviors in thermodynamic state functions are observed also for these domains. Dotted curves in the subfigures of Fig. 2 are the predictions of our analytical model, which will be examined in the following section. The physical mechanism of the decrement in free energy in spite of a decrease in entropy is also investigated and explained in detail. Similar investigation is done for local increment in entropy (Fig. 2f) observed for a certain angular interval.

\text{ }\\
\noindent\textbf{Overlapped quantum boundary layer method: Analytical predictions and physical explanations of QShE.}
So far we showed the shape dependence of free energy, entropy and internal energy based on numerical calculations. In order to explore the underlying physical mechanisms of QShE as well as to predict them analytically, we invoke the quantum boundary layer (QBL) concept\cite{qbl}. Particles confined in nanoscale domains cannot get arbitrarily close to the impenetrable boundaries of the domain due to their wave nature. As a consequence of ensemble average of quantum probability density, particle density goes to zero near to the boundaries and particles occupy smaller volume (effective volume) than the actual one \cite{qbl,uqbl,nanocav,pdx,qforce}. This inhomogeneous density region is called QBL. Non-uniform density distribution can be approximated by a uniform one by assuming an empty layer (excluded region \textit{i.e.} zero density region) near boundaries and a flat distribution at the rest of the domain (see Fig. 10 in Methods section). Since inhomogeneous parts of density distribution are approximated by step functions, it is called zeroth order QBL approach and it allows to define effective volumes easily for any domain shape. For Maxwell-Boltzmann gases, thickness of this QBL is obtained as $\delta=L_c/2\sqrt{\pi}$ which is in the order of thermal de Broglie wavelength of particles. In literature, QBL method has been first used to obtain QSE terms for thermodynamic properties directly from their conventional expressions without solving Schr\"odinger equation and using PSF or Weyl conjecture \cite{qbl,nanocav}. Different methods to calculate QSE on thermodynamic properties are briefly discussed in Methods section. Analytical procedure for the calculation of QShE is built on top of the QSE calculation methodology based on QBL method. That's why we frequently refer to QSE procedures in this section.

Classically, density distribution of a gas inside a domain is uniform, $n_{cl}=N/V$, at thermodynamic equilibrium. On the other hand, the true density is calculated by the ensemble average of quantum probability density as $n(\textbf{r})=\left\langle\left|\Psi(\textbf{r})\right|^2\right\rangle_{\mli{ens}}=\sum_\varepsilon{\exp(-\varepsilon\beta)}\left|\Psi_{\varepsilon}(\textbf{r})\right|^2/\sum_\varepsilon{\exp(-\varepsilon\beta)}$ where $\left|\Psi(\textbf{r})\right|^2$ is the probability density and summations include possible degeneracies if any. In Fig. 3, normalized density distributions, $\tilde{n}=n/n_{cl}$, of particles confined in considered domains are shown for four different rotation angles. As is seen, the existence of QBLs is clear and density distributions of particles confined in nanoscale domains are non-uniform even at thermodynamic equilibrium due to their quantum nature \cite{qbl,uqbl,nanocav,pdx,qforce}.

\begin{figure}[t]
\centering
\includegraphics[width=0.49\textwidth]{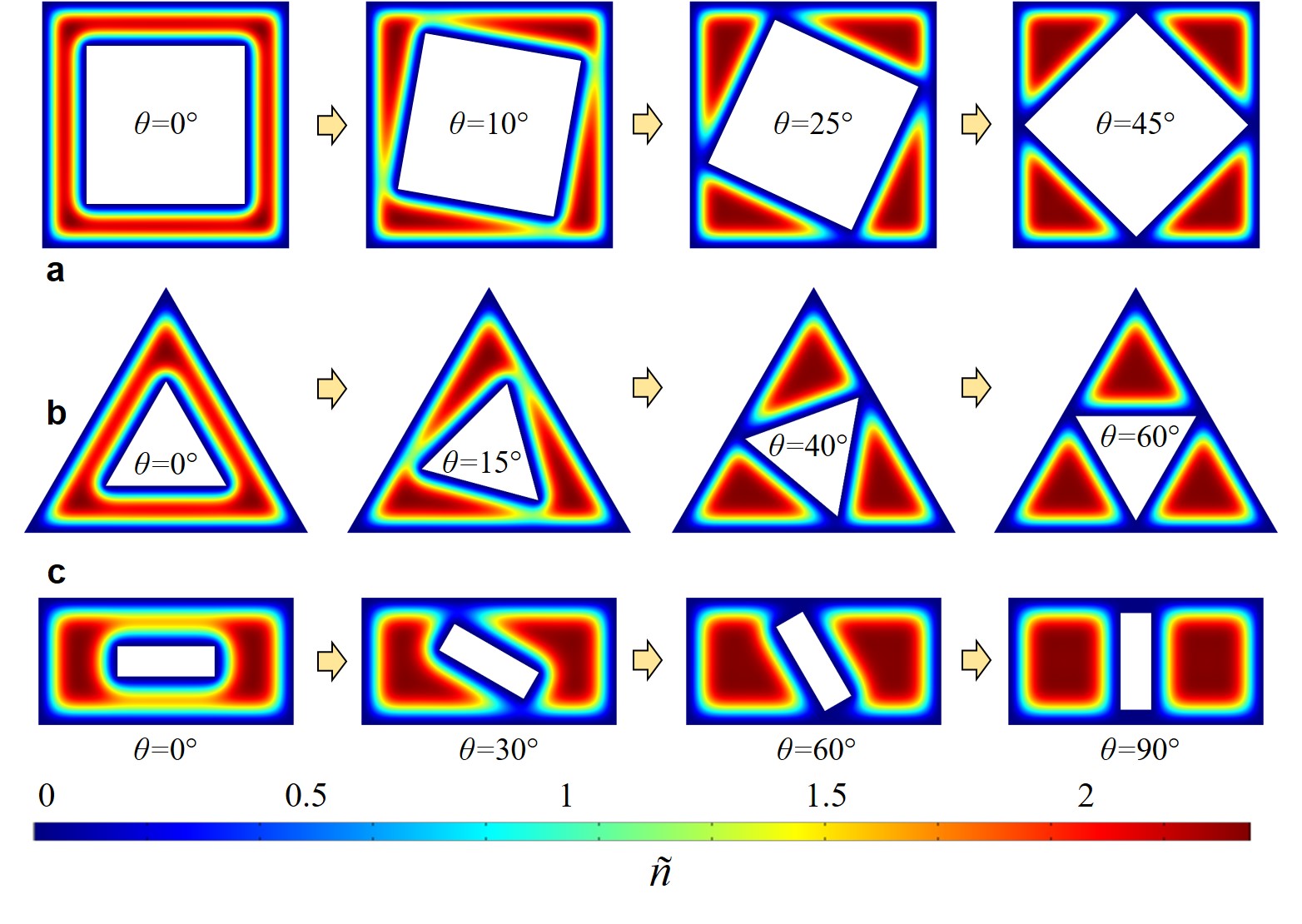}
\caption{\bfseries{Density profiles of particles.} \mdseries Normalized ($\tilde{n}=n(\textbf{r})/n_{cl}$) density distributions of particles confined in nested (\textbf{a}) square, (\textbf{b}) triangle, (\textbf{c}) rectangle domains are presented for four different angles of rotation. As a result of their wave nature, particles have tendency to stay away from boundaries and accumulate into less confined regions of the domain creating a non-uniform density distribution even at thermodynamic equilibrium.}
\label{fig:pic3}
\end{figure}

Note that in Fig. 3 particles tend to stay away from boundaries or narrow regions and occupy smaller volume than the geometric volume (apparent volume). Using this effective volume occupied by particles it's possible to predict QSE directly from classical expressions. Replacing the actual size parameter (\textit{e.g.} $V$, geometric volume in 3D) in classical partition function expression with the effective size parameter (\textit{e.g.} $V_{\mli{eff}}^{\mli{novr}}$) generates QSE correction terms and provides a physical explanation with a better understanding of QSE on thermodynamic behaviors as well\cite{qbl}. Effective volume is simply determined by $V_{\mli{eff}}^{\mli{novr}}=V-V_{\mli{qbl}}$ where $V_{\mli{qbl}}$ is the excluded volume as a result of QBLs. The superscript $novr$ indicates the condition in which QBLs do not overlap on each other. By this way, QBL method gives exactly the same QSE terms that Weyl conjecture or PSF gives and it serves as a concrete and simple way to predict QSE in confined systems. However, this procedure cannot predict the shape dependence in its current form since $V_{\mli{eff}}^{\mli{novr}}$ does not depend on any shape related parameter (here $\theta$).

Calculation of the volume of excluded regions defined by QBLs, however, has to be done with caution in strongly confined domains. When we carefully apply the QBL method on the nested square domain for instance, we see that QBLs of inner and outer domain boundaries overlap with each other at some angular configurations (Fig. 4). This overlap volumes actually carry crucial information about the shape of the domain, as the amount of overlap is precisely depend on the shape ($\theta$). In the usual QBL method, excluded volume ($V_{\mli{qbl}}$) is subtracted from the geometric one to find the effective volume. However, in case of overlap, volumes of overlapped QBL regions are considered twice improperly during the subtraction process. To obtain the true effective volume, this overlap volume has to be added to $V_{\mli{eff}}^{\mli{novr}}$, to correct the subtraction process. By using QBL method and carefully dealing with overlap regions, the true effective volume is given as
\begin{equation}
\begin{split}
V_{\mli{eff}}&=V-V_{\mli{qbl}}+V_{\mli{ovr}}=V_{\mli{eff}}^{\mli{novr}}+V_{\mli{ovr}}\\
&=V-\delta A+\delta^2 P-\delta^3 N_V+V_{\mli{ovr}}.
\end{split}
\end{equation}
Here, $V$ is geometric volume (classical term), $V_{\mli{qbl}}$ represents the excluded volume determined by QBLs (leads to QSE corrections), $V_{\mli{eff}}^{\mli{novr}}=V-V_{\mli{qbl}}$ is QSE corrected volume and $V_{\mli{ovr}}$ is the overlap volume determined by the amount of overlapped QBLs (leads to QShE corrections). $\theta$ dependency of $V_{\mli{ovr}}$ is analytically given by Eq. (10) in Methods section.

\begin{figure}[t]
\centering
\includegraphics[width=0.36\textwidth]{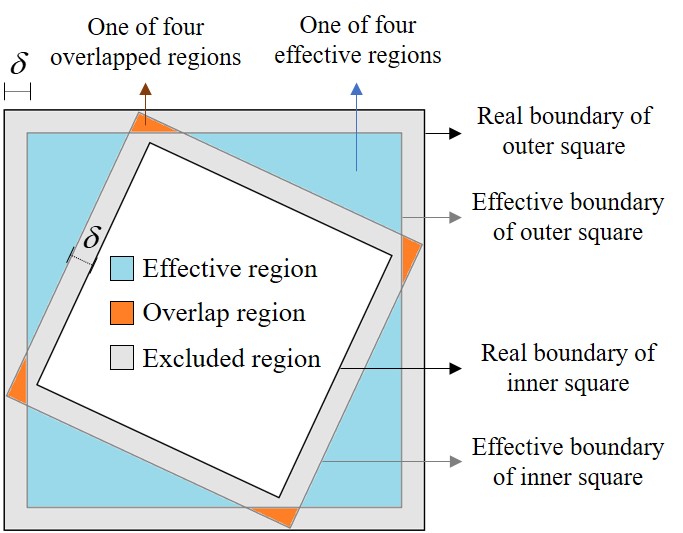}
\caption{\bfseries{Overlapped QBL method: Effective, overlap and excluded regions.} \mdseries Effective, overlap and excluded regions due to QBL inside the nested square domain is shown. QBLs of outer and inner squares overlap with each other in certain rotational configurations and lead to the miscalculation of effective regions in the usual QBL method which corresponds to Weyl conjecture. Considering the influence of overlap regions not only corrects the effective volume calculation, but also predicts the shape dependence quite accurately.}
\label{fig:pic4}
\end{figure}

As is seen from Fig. 5, the true effective volumes are shape dependent, because the amounts of overlap volumes are different in each degree configuration. While there is no overlap at small degrees according to 0th order QBL approach, overlaps of QBLs start after a critical angle value (denoted by $\theta_{*}$ for nested square domain for which the analytical expressions are explicitly derived in Methods section). Therefore, effective volume increases as $\theta$ goes from $\theta_{*}$ till $45^{\circ}$ where effective volume reaches to its maximum and free energy goes to its minimum. This behavior reveals the profound link between thermodynamics and geometry; minimization of Helmholtz free energy under isothermal process is equivalent to the maximization of effective volume rather than the geometric volume.

\begin{figure}[t]
\centering
\includegraphics[width=0.49\textwidth]{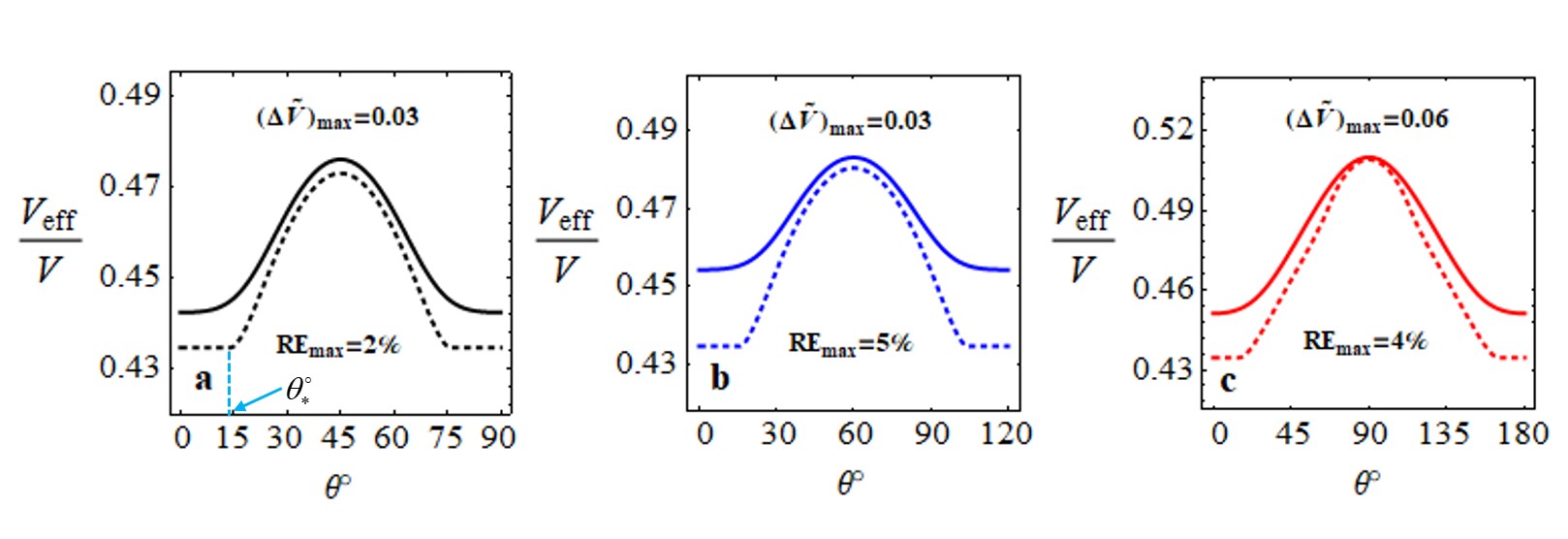}
\caption{\bfseries{Variations of effective volumes with rotation angle.} \mdseries Variations of effective to geometric volume ratio with $\theta$ for (\textbf{a}) nested square, (\textbf{b}) nested triangle and (\textbf{c}) nested rectangle domains. Solid and dotted curves represent numerical and analytical solutions respectively. $(\Delta\tilde{V})_{\mli{max}}$ gives the maximum change in effective to geometric volume ratio for numerical cases and $RE_{\mli{max}}$ indicates the maximum relative errors of analytical predictions at each case. In all cases, the domain is so confined that excluded volume is larger than effective volume.}
\label{fig:pic5}
\end{figure}

Owing to the logarithmic dependence of Helmholtz free energy on effective volume, it is possible to separate QSE and QShE contributions and write any thermodynamic state function as the addition of classical, QSE and QShE terms, $X=X_{cl}+X_{\mli{QSE}}+X_{\mli{QShE}}$. Thus, we can obtain the analytical expression for size and shape dependent Helmholtz free energy as
\begin{equation}
\begin{split}
\tilde{F}=\frac{F}{Nk_BT}= &\left(-\ln\frac{V}{N(4\delta)^3}-1\right)_{\mkern-7mu\mli{cl}}+\left[-\ln\left(1-\frac{V_{\mli{qbl}}}{V}\right)\right]_{\mkern-2mu\mli{QSE}} \\
&+\left[-\ln\left(1+\frac{V_{\mli{ovr}}}{V_{\mli{eff}}^{\mli{novr}}}\right)\right]_{\mkern-2mu\mli{QShE}},
\end{split}
\end{equation}
where the first term denoted by bracket subscript $cl$ is the classical (bulk) term that is the result of integral representation of summations under continuum approximation. The second term represents QSE corrections based on bounded continuum approximation which can be predicted by PSF, Weyl conjecture or usual QBL methods equivalently. The last term is the term responsible for QShE, which cannot be determined by any of these methods, but overlapped QBL method. It seems that these three terms correspond to three terms of PSF: integral, zero correction and discrete correction terms respectively \cite{discnat}. Likewise, we get dimensionless entropy as
\begin{equation}
\begin{split}
\tilde{S}=\frac{S}{Nk_B}= &\left(\ln\frac{V}{N(4\delta)^3}+\frac{5}{2}\right)_{\mkern-7mu\mli{cl}} \\
&+\left[\ln\left(1-\frac{V_{\mli{qbl}}}{V}\right)-\frac{T}{V_{\mli{eff}}^{\mli{novr}}}\frac{\partial V_{\mli{qbl}}}{\partial T}\right]_{\mkern-2mu\mli{QSE}} \\
&+\Bigg[\underbrace{\ln\left(1+\frac{V_{\mli{ovr}}}{V_{\mli{eff}}^{\mli{novr}}}\right)}_{S_{\mli{QShE}}^{\mli{I}}}+\underbrace{\frac{V_{\mli{eff}}^{\mli{novr}} T}{V_{\mli{eff}}}\frac{\partial}{\partial T}\frac{V_{\mli{ovr}}}{V_{\mli{eff}}^{\mli{novr}}}}_{S_{\mli{QShE}}^{\mli{II}}}\Bigg]_{\mkern-2mu\mli{QShE}}
\end{split}
\end{equation}
and internal energy is then given by
\begin{equation}
\begin{split}
\tilde{U}=\frac{U}{Nk_BT}=& \left(\frac{3}{2}\right)_{\mkern-7mu\mli{cl}}+\left(-\frac{T}{V_{\mli{eff}}^{\mli{novr}}}\frac{\partial V_{\mli{qbl}}}{\partial T}\right)_{\mkern-7mu\mli{QSE}} \\
&+\left(\frac{V_{\mli{eff}}^{\mli{novr}} T}{V_{\mli{eff}}}\frac{\partial}{\partial T}\frac{V_{\mli{ovr}}}{V_{\mli{eff}}^{\mli{novr}}}\right)_{\mkern-7mu\mli{QShE}}.
\end{split}
\end{equation}
Evidently, QShE contributions to thermodynamic state functions are described by overlap volumes.

Details of the analytical expressions of shape dependent thermodynamic state functions are given in the Methods section of this article. Comparison of analytical results (based on overlap volumes) with numerical ones (based on the calculation of summations in the definitions of thermodynamic state functions using energy eigenvalues from the numerical solution of Schr\"odinger equation) are shown in Fig. 2 where solid and dotted curves are the results of numerical and analytical calculations (Eqs. 3, 4 and 5) respectively. When overlap volumes are considered, functional behaviors of thermodynamic quantities under QShE are correctly predicted by analytical expressions with reasonable errors for various confinement domains. For nested square, triangle and rectangle domains, overlap volumes become maximum at $45^{\circ}$, $60^{\circ}$ and $90^{\circ}$ respectively (Fig. 5) and minimum difference between analytical and numerical predictions are also achieved for these configurations (Fig. 2). The zeroth order QBL approach based on homogenized density distribution assumes empty regions with thickness of $\delta$ near to boundaries and so couldn't account weak overlaps which may actually occur in inhomogeneous parts (approximately thickness of $2\delta$) of density distributions (Fig. 10). That's why, for certain values of rotation angles corresponding to non-overlap conditions ($\theta<\theta_{*}$) in the zeroth order QBL model, the difference between numerical and analytical results reach to their maximum (Figs. 2 and 5).

There is a fundamental difference between the mechanisms of QSE and QShE. While keeping the geometric volume constant, if we increase the other size parameters, effective volume decreases, so confinement increases due to the existence of QBL. This type of confinement is a size-dependent confinement, shortly size-confinement. The larger the QBL, the smaller the effective volume and the stronger QSE. Unlike in QSE, while keeping all size parameters constant, if we just change shape by $\theta$ to decrease the overlap volume, effective volume decreases, so confinement increases. This type of confinement is shortly called shape-confinement. The larger the overlap volume, the larger the effective volume and the stronger QShE. Thus, QSE and QShE work opposite in terms of confinement. The former decreases effective volume, while the latter increases it. Although a strong confinement is needed to reveal QShE, its presence actually functions against to confinement. As the shape-confinement decreases from $\theta=0^{\circ}$ to $\theta=45^{\circ}$ (shape-deconfinement), confinement energy also decreases which explains internal energy decrease shown in Figs. 2g, 2h and 2i.

\text{ }\\
\noindent\textbf{Unusual behavior of entropy under QShE.}
The most distinctive feature of QShE appears in the behavior of entropy under shape variation. As is seen from Fig. 2 entropy decreases simultaneously with free energy and internal energy. This is an uncommon behavior as Helmholtz free energy and entropy behave oppositely in most of the systems. For example, when we compare the isothermal shape-deconfinement from $\theta=0^{\circ}$ to $\theta=45^{\circ}$, with an ordinary isothermal expansion, in the latter, the geometric volume increases due to the expansion of the system, whereas in the former, geometric volume does not change but the effective volume increases. Both in isothermal geometric and effective  volume expansions; confinement of the domain, energy eigenvalues, free energy and internal energy decrease. On the other hand, while entropy always increase in geometric expansion, it can either increase (Fig. 2f around $\theta=90^{\circ}$) or even decrease (Figs. 2d, 2e and 2f) during isothermal effective volume expansion (shape-deconfinement) process. Decrement of entropy during an isothermal effective volume expansion is an unexpected behavior and against the classical intuition, since volume expansion always increases entropy classically.

In order to understand this unusual behavior due to QShE, we focus on how QShE influence thermodynamic quantities. Variations of QShE terms of free energy, entropy and internal energy (see Eqs. 3, 4 and 5) with overlap volume percentages, by considering their ratios to the classical expressions of the relevant quantities are given in Fig. 6. The higher the overlap volumes means the larger the QShE. Columns from left to right in Fig. 6 indicate the relevant thermodynamic properties for nested square, triangle and rectangle domains. Overall, QShE cause decrements in free energy, entropy and internal energy in general unlike QSE.

\begin{figure}[b]
\centering
\includegraphics[width=0.45\textwidth]{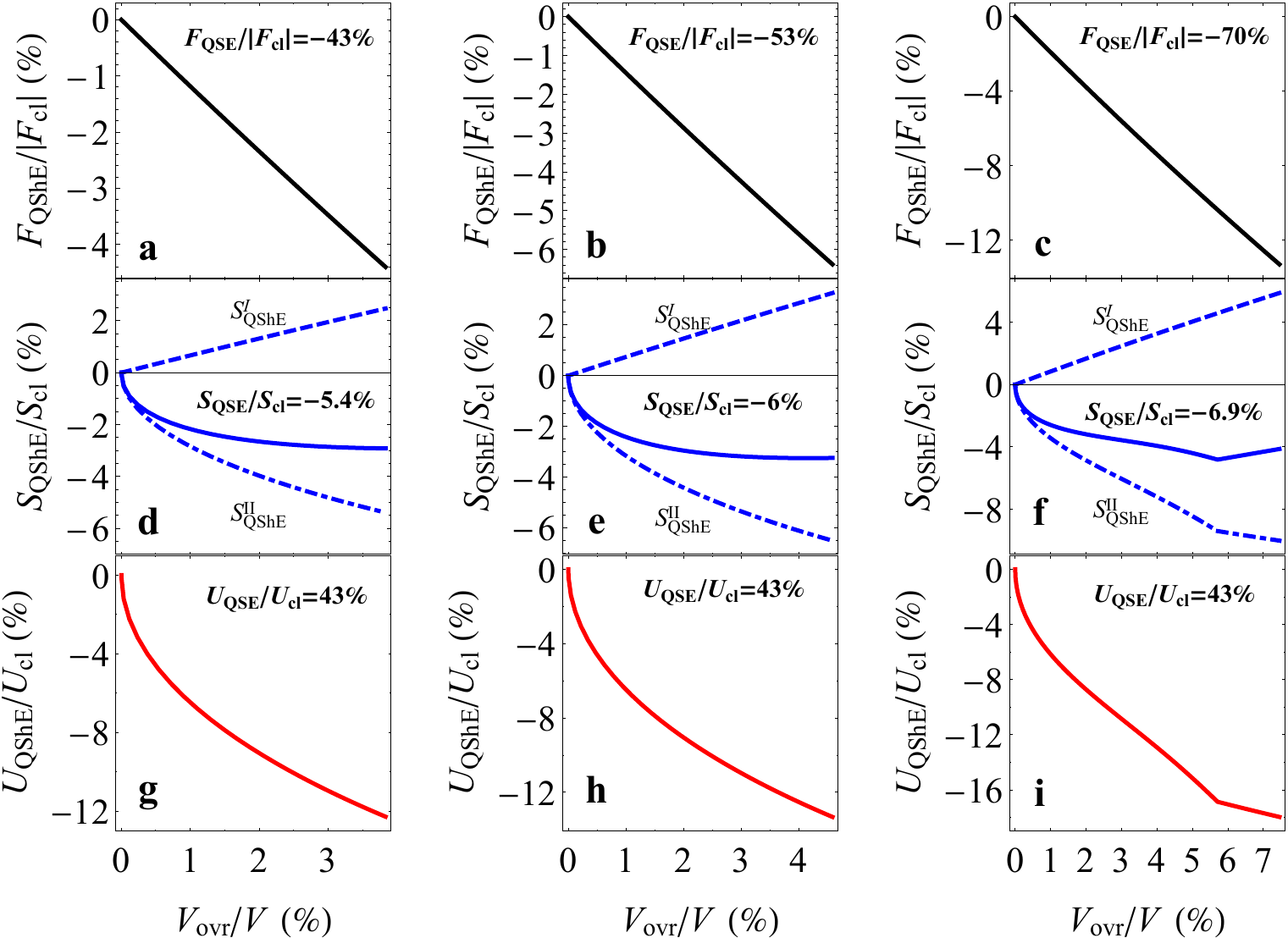}
\caption{\bfseries{Overlap volume dependence of QShE terms of thermodynamic state functions.} \mdseries Variations of normalized shape-dependent terms (obtained from the analytical expressions) of free energy, entropy and internal energy with overlap volume percentages are presented by black, blue and red curves respectively. Dashed blue and dot-dashed blue curves represent the first and second terms of shape-dependent terms of entropy. First, second and third columns represent the examination of the related quantities for nested square, triangle and rectangle domains respectively. Percentages of QSE terms' contributions are also given inside each subfigure for comparison.}
\label{fig:pic6}
\end{figure}

For the QSE term in free energy under weak confinement regimes, $V_{\mli{qbl}}<<V$ condition is satisfied and since $\ln(1+x)\approx x$ for $x<<1$, a linear behavior occurs\cite{qbl,nanocav}. A similar linear behavior appear in QShE term of free energy expression as long as $V_{\mli{ovr}}<<V_{\mli{eff}}^{\mli{novr}}=V-V_{\mli{ovr}}$ condition is valid. That's why, QShE term of free energy decreases linearly for small values of overlap volume ratio (Fig. 6a, 6b and 6c). Nevertheless, linear approximation has never been used in this article.

Two shape-dependent terms of entropy, $S_{\mli{QShE}}^{\mli{I}}$ and $S_{\mli{QShE}}^{\mli{II}}$ in Eq. (4), compete with each other on the determination of overall behavior (which is also the case in entropy's QSE terms). The first term is the same shape-dependent free energy term ($F_{\mli{QShE}}$) except with opposite sign and so its contribution to entropy is always positive. Therefore, this term causes the usual behaviors in entropy and free energy. In other words, entropy increases and free energy decreases while the effective volume increases due to increment in overlap volume. On the contrary, the second term contribute to entropy negatively. $V_{\mli{ovr}}$ is inversely proportional with temperature since increasing temperature reduces the thickness of QBLs and cause overlap volumes to diminish. That's why the contribution of $S_{\mli{QShE}}^{\mli{II}}$ term is negative. The unusual behavior of entropy directly comes from the second term which actually depends on the temperature sensitivity of overlap volume. This sensitivity is negatively proportional to the square root of overlap volume itself dominantly (see Eq. (11) in Methods section). This is the reason why entropy decreases while overlap volume increases. In other words, increment in overlap volume affects entropy in two different competing ways: one is the entropy increment due to increment of effective volume, the other is entropy decrement due to negative temperature sensitivity of effective volume. Overall behavior, however, depends on the result of the competition of these two terms. That's why a special attention need to be given to the behavior of entropy and internal energy in nested rectangle domain, see Figs. 6f and 6i. Unlike to nested square and triangle cases, for some rotational configurations at nested rectangle case, $S_{\mli{QShE}}^{\mli{I}}$ term starts to dominate the behavior of entropy and leads to its increment. Therefore, depending on the shape of the domain, different QShE terms determine the behavior of entropy. The term responsible from the decrease in entropy ($S_{\mli{QShE}}^{\mli{II}}$) also decreases internal energy. Therefore, absolute value of $U_{\mli{QShE}}/U_{cl}$ has a square root relationship with overlap volume percentages.

\text{ }\\
\noindent\textbf{A torque induced by QShE.}
Let's recall the variation of free energy with $\theta$ for the nested square domain. It's seen that the lowest free energy state (thermodynamically stable one) is $\theta=45^{\circ}$ state. Any other angular configuration is unstable and the system tends to turn into the lowest free energy configuration eventually. If the initial state of the system is different than $45^{\circ}$, the inner wire spontaneously starts to rotate until it reaches to $45^{\circ}$ configuration.

In order for inner wire to rotate, there has to be a torque exerted on the inner wire by confined particles. It should equal to the minus of the derivative of free energy with respect to rotation angle $\theta$.
Then, the torque can numerically be calculated for any $\theta$ as follows,
\begin{equation}
\mathcal{T}=-\frac{F_{\theta+\Delta\theta}-F_{\theta}}{\Delta\theta}=-\frac{Nk_BT}{\Delta\theta}\ln\left(\frac{\zeta_\theta}{\zeta_{\theta+\Delta\theta}}\right),
\end{equation}
where $\theta$ and $\theta+\Delta\theta$ subscripts denote initial and perturbed rotational states respectively. $\Delta\theta$ denotes the amount of the rotational perturbation, which has to be very small (here it is choosen as $\Delta\theta$=0.1). The variation of torque with rotation angle is shown in Fig 7a where solid black and dashed red curves denote the results of numerical and analytical calculations respectively. The analytical result is obtained from the derivative of Eq. (3) with respect to $\theta$ by considering Eq. (10). The torque becomes maximum at $\theta=26^{\circ}$ configuration and vanishes at $\theta=0^{\circ}$ and $\theta=45^{\circ}$ in which density distribution of particles has axial symmetry inside the domain.

\begin{figure}[t]
\centering
\includegraphics[width=0.45\textwidth]{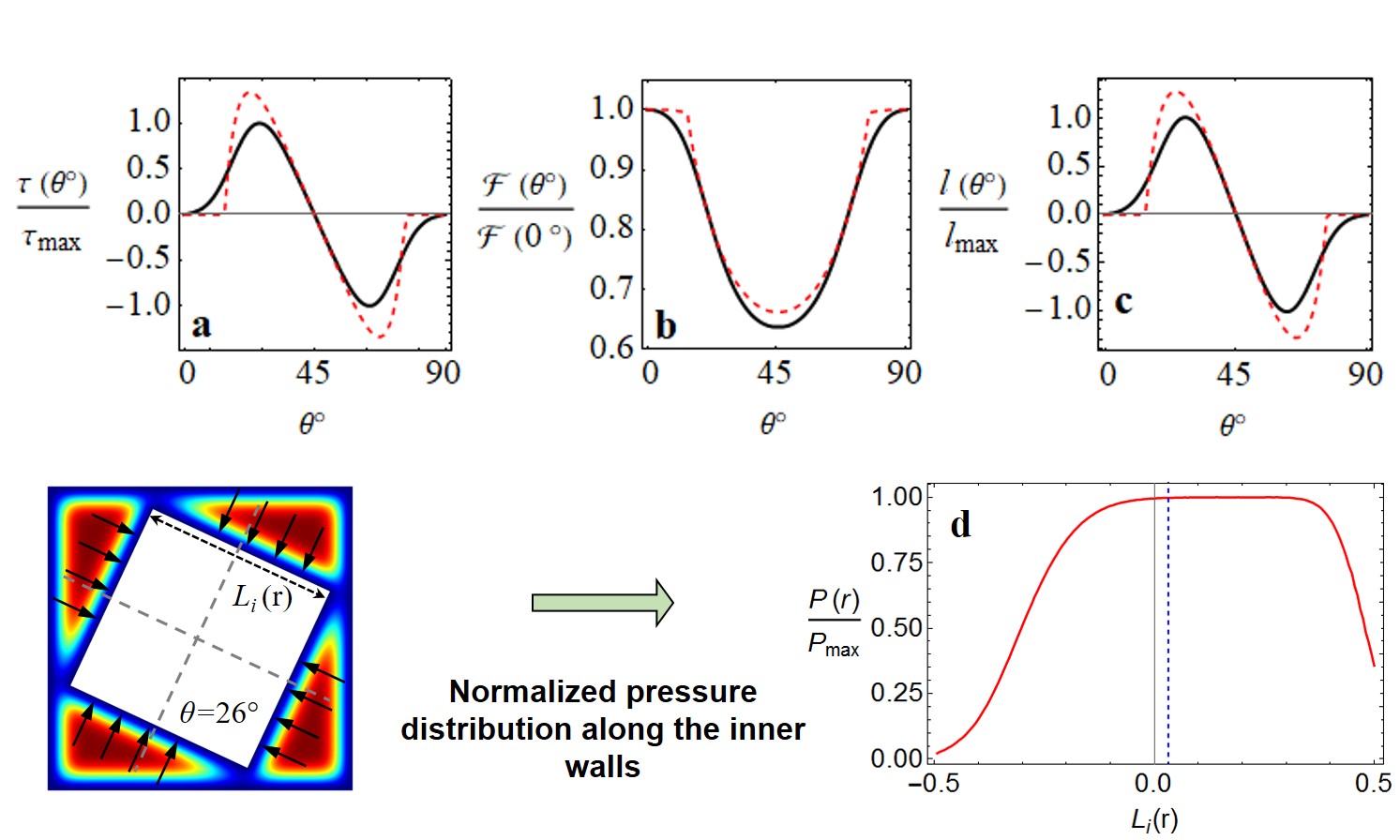}
\caption{\bfseries{Non-uniform pressure distribution causes a torque.} \mdseries A torque is induced as a result of non-uniform pressure distribution caused by inhomogenous density distribution due to wave character of particles. Solid black and dashed red curves in \textbf{a}, \textbf{b} and \textbf{c} represent the results of numerical and analytical calculations respectively. (\textbf{a}) Configurational ($\theta$) dependency of the torque. For the sake of comparability $\tau_{max}$ obtained from numerical calculations is used. Torque reaches to its maximum value at $26^{\circ}$. (\textbf{b}) Total pressure force on the walls of inner square wire and (\textbf{c}) its application point varying with $\theta$. (\textbf{d}) For $\theta=26^{\circ}$ configuration, pressure distribution along one of the walls of inner square is shown by solid red curve. The pressure distribution at $26^{\circ}$ is normalized by dividing all values to the maximum one. Blue dotted vertical line denotes the location of the application point of total pressure force.}
\label{fig:pic7}
\end{figure}

We can also calculate total amount of pressure forces exerted on the walls of inner square by positively perturbing the size of inner square wall by a tiny amount of $\Delta h$ and probing the mechanical response of the system to this perturbation over Helmholtz free energy, from the following equation
\begin{equation}
\mathcal{F}=-\frac{F_{L_i+\Delta h}-F_{L_i}}{\Delta h}=-\frac{Nk_BT}{\Delta h}\ln\left(\frac{\zeta_{L_i}}{\zeta_{L_i+\Delta h}}\right),
\end{equation}
where $L_i$ and $L_i+\Delta h$ subscripts denote initial and perturbed size of inner square respectively for a certain $\theta$ degree configuration. As it is seen from Fig 7b, total pressure forces decrease from $0^{\circ}$ to $45^{\circ}$, which supports the decrease in effective confinement of the domain. Application points of pressure forces can easily be found by dividing Eq. (6) to Eq. (7), Fig. 7c. 

This torque actually originates from a non-uniform pressure distribution along the walls of inner wire. In classical picture, density and so the pressure distributions are homogeneous and there is no torque at thermodynamic equilibrium. In quantum picture, however, due to the wave nature of particles, local density distribution becomes inhomogeneous \cite{qbl} and this cause non-uniform pressure distribution along the walls even at thermodynamic equilibrium. As is seen from Fig. 3a, density profiles of particles in each quadrant are symmetric for $\theta=0^{\circ}$ and $\theta=45^{\circ}$ configurations, whereas for any degree configuration in between (like $\theta=10^{\circ}$ or $\theta=25^{\circ}$), particles escape from relatively narrow regions to wider regions due to QBL and cause axially asymmetric pressure distribution on the walls of inner square.

The straightest way to investigate the pressure distribution along the walls of inner square is to make tiny inward perturbations along the wall and calculate the free energy difference (boundary deformation work) of the system as a response to these local perturbations. With this approach, we are able to numerically prob the local pressure along the walls of inner square and find its distribution. The results of the numerical calculations show the non-uniformity of the pressure in Fig. 7d. Energy eigenvalues of particles are calculated by numerically solving the Schr\"odinger equation and used to calculate free energy repeatedly for each perturbative case. The details of the calculations are given in Methods section. The obtained pressure distribution is apparently right-skewed (Fig. 7d), which causes to a shift in the application point of total pressure force over the surface and justifies the existence of the torque. Therefore, the same application point can also be obtained by using the local pressure distributions, $P(r)$, Fig. 7d. In this case, the application point of pressure forces is simply given by $l=\int_{-L_i/2}^{L_i/2}rP(r)dr/\int_{-L_i/2}^{L_i/2}P(r)dr$, (Fig. 7d), which shows the consistency of different approaches. As expected, dependence of application points on rotation angle (Fig. 7c) mimics the behavior of torque with respect to $\theta$.

The torque is linearly proportional with particle number inside the system. While pressure forces increase with increasing temperature, torque decreases because of a larger contraction in application points of pressure forces due to decrease in QBL. The torque is very sensitive to the changes in the sizes of the confinement domain. Increasing the sizes of both square wires simultaneously or increasing the ratio of outer and inner square sizes decreases overlap volume and so decreases torque extensively.

\text{ }\\
\noindent\textbf{Isoformal process and a new type of thermodynamic cycle.}
Inclusion of shape as a control parameter in thermodynamic state functions allows us to define a new type of thermodynamic process called isoformal process where the shape of the system is kept constant. Thus, QShE give rise to thermodynamic cycles that never proposed before. In this section, we introduce a novel thermodynamic cycle consisting of two isoformal and two isothermal processes. The cycle consists of four consecutive steps as shown in Fig. 8: (\textbf{1}) \textit{Isoformal heat addition} $1\rightarrow 2$: The temperature of the system increases from $T=200$K to $T=300$K. (\textbf{2}) \textit{Isothermal shape-confinement} $2\rightarrow 3$: Inner wire is rotated from $45^{\circ}$ to $0^{\circ}$ position by performing work on it. At this process heat is also given to the system to keep the temperature constant at $300$K, which is against to usual expectations. (\textbf{3}) \textit{Isoformal heat rejection} $3\rightarrow 4$: The temperature of the system decreases from $T=300$K back to $T=200$K, while keeping the inner wire at the $0^{\circ}$ position. (\textbf{4}) \textit{Isothermal shape-deconfinement} $4\rightarrow 1$: Inner wire rotates from $0^{\circ}$ to $45^{\circ}$ degree position by doing work and rejecting heat by keeping the temperature constant. All processes are assumed to be infinitely slow (quasi-equilibrium), hence reversible, and there is no coherence among energy levels.

\begin{figure}[t]
\centering
\includegraphics[width=0.49\textwidth]{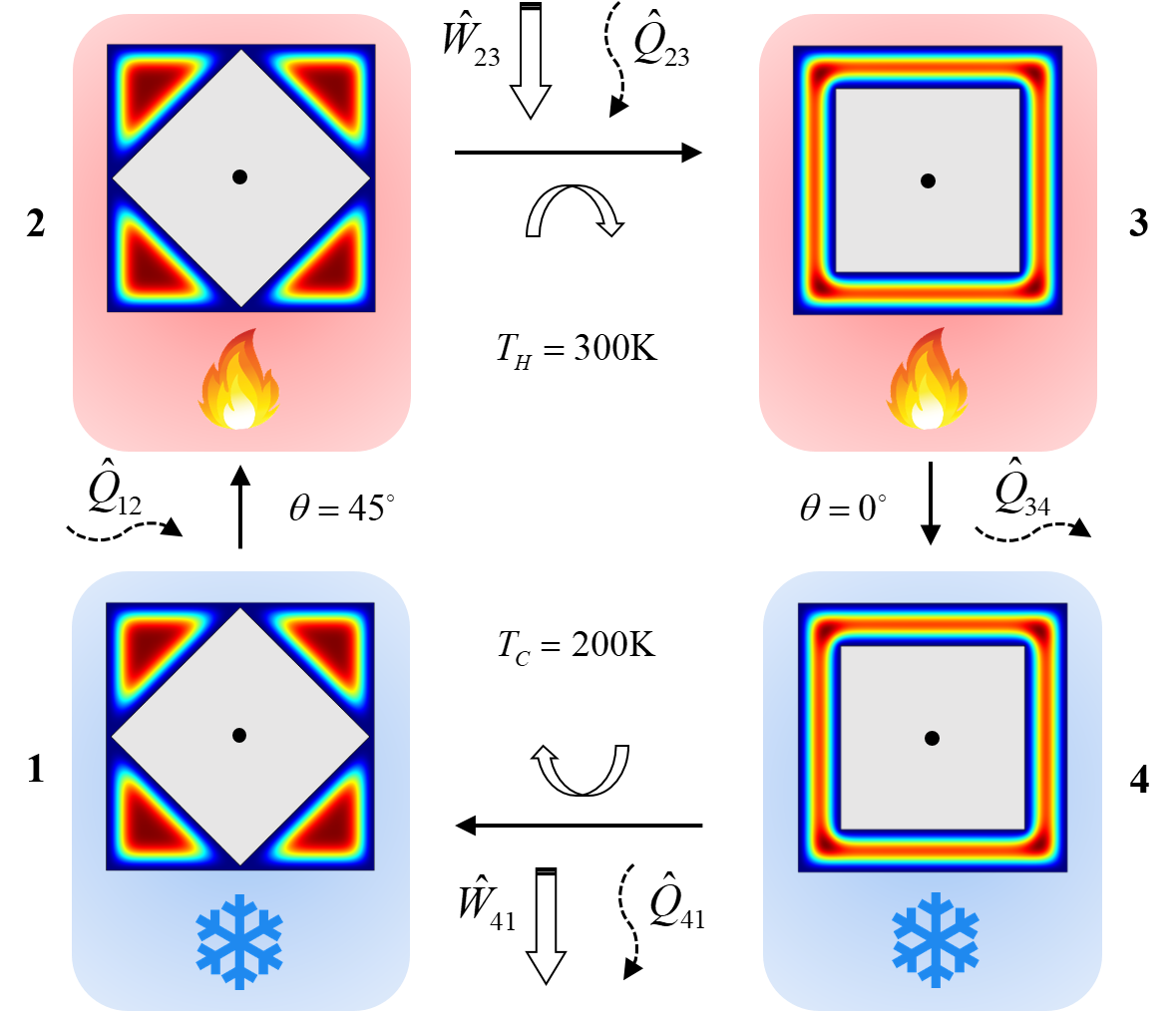}
\caption{\bfseries{A novel thermodynamic cycle.} \mdseries The proposed thermodynamic cycle based on QShE. The cycle consists of four processes: (\textbf{1}) isoformal (shape preserving) heat addition, (\textbf{2}) isothermal shape-confinement, (\textbf{3}) isoformal heat rejection, (\textbf{4}) isothermal shape-deconfinement. The cycle generates work during isothermal shape-deconfinement process.}
\label{fig:pic8}
\end{figure}

In Fig. 9, $T$-$S$ and $\tau$-$\theta$ diagrams of the cycle are shown. On the contrary to classical expectations, work exchange during high temperature shape transformation is less than that of low temperature one, Fig. 9b. Also, work and heat exchanges under isothermal shape transformation processes are in the same direction unlike the ones in isothermal volume variation, Fig. 9a and 9b. It should be noted that, it is not always necessary to mechanically rotate the inner wire in a closed system to be able to realize $2\rightarrow 3$ and $4\rightarrow 1$ processes. Instead, a gas flow can also be considered in a channel made by nested wires where the inner one is twisted from $45^{\circ}$ to $0^{\circ}$ and from $0^{\circ}$ to $45^{\circ}$ along the channel. Therefore, an electron gas in semiconductors or conductors can be used to realize this cycle. 

For this cycle, heat and work exchanges at each process are simply determined as follows
\begin{equation}
\begin{split}
Q_{12}=& U\left(T_H,45^{\circ}\right)-U\left(T_C,45^{\circ}\right) \\
Q_{23}=& T_H\left[S\left(T_H,0^{\circ}\right)-S\left(T_H,45^{\circ}\right)\right] \\
W_{23}=& \int_{45}^{0}{\tau\left(T_H,\theta\right)d\theta} \\
=& U\left(T_H,0^{\circ}\right)-U\left(T_H,45^{\circ}\right)-Q_{23} \\
Q_{34}=& U\left(T_C,0^{\circ}\right)-U\left(T_H,0^{\circ}\right) \\
Q_{41}=& T_C\left[S\left(T_C,45^{\circ}\right)-S\left(T_C,0^{\circ}\right)\right] \\
W_{41}=& \int_0^{45}{\tau\left(T_C,\theta\right)d\theta} \\
=& U\left(T_C,45^{\circ}\right)-U\left(T_C,0^{\circ}\right)-Q_{41}.
\end{split}
\end{equation}

\begin{figure}[t]
\centering
\includegraphics[width=0.45\textwidth]{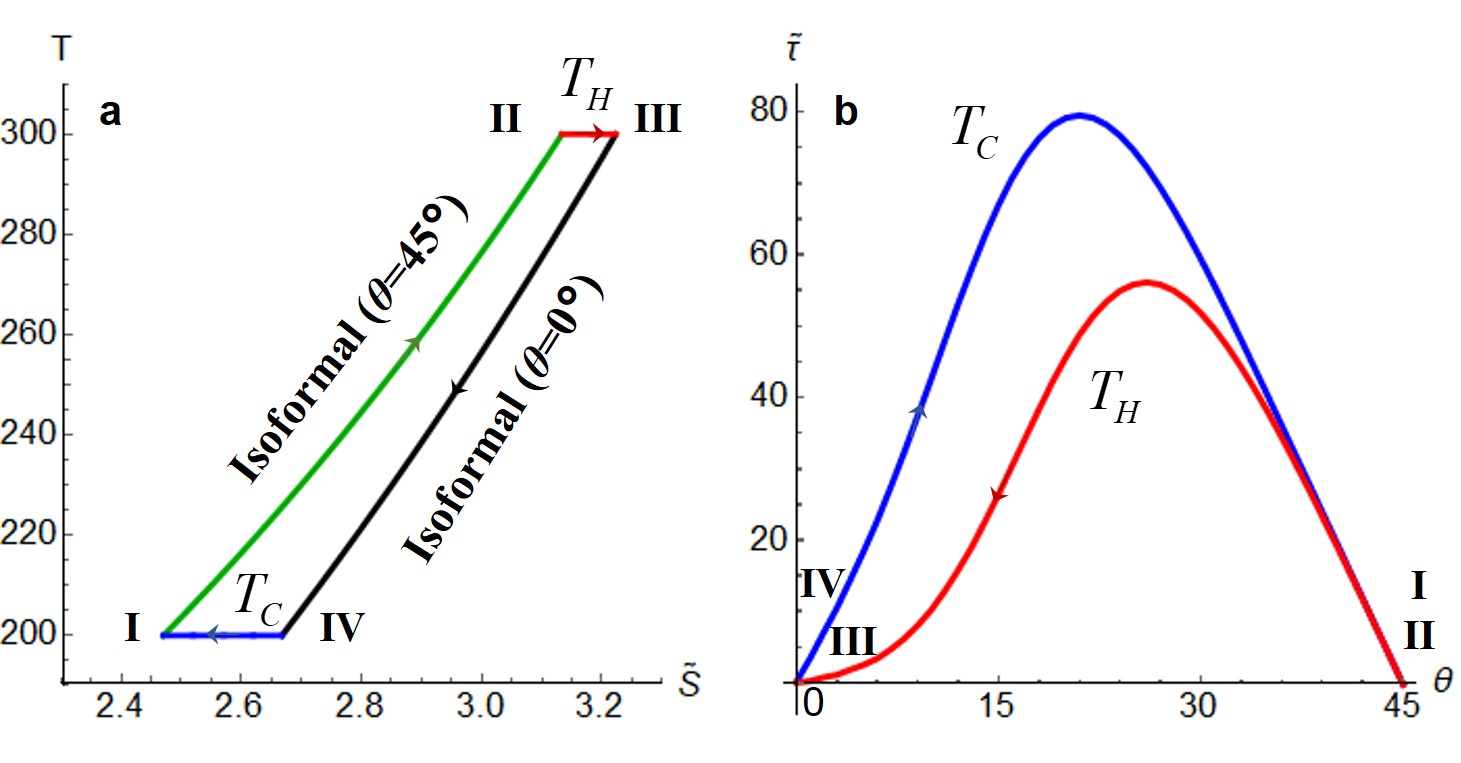}
\caption{\bfseries{$T$-$S$ and $\tau$-$\theta$ diagrams.} \mdseries (\textbf{a}) $T$-$S$ diagram. (\textbf{b}) $\tau$-$\theta$ diagram. From state \textit{1} to \textit{2} and \textit{3} to \textit{4}, no change happens in $\tau$-$\theta$ diagram since $\theta$ is kept constant during these change of states. Also, the torque vanishes at $\theta=0^{\circ}$ and $\theta=45^{\circ}$ configurations. As the magnitude of QShE increase with decreasing temperature, higher amount of torque occurs at lower temperatures on the contrary to classical expectations.}
\label{fig:pic9}
\end{figure}

Using Eq. (8), heat input and net work output are determined respectively as $Q_{in}=Q_{23}+(Q_{12}-Q_{34})$ and $W_{\mli{net}}=W_{23}+W_{41}$. For the sizes and density given in the Methods section, the cycle efficiency is obtained as $25\%$ for $T_H=300$K and $T_C=200$K. A refrigeration cycle can definitely be obtained just by reversing the power cycle presented in Fig. 8. By considering QShE on thermodynamic properties, it is possible to define many new thermodynamic processes and cycles like the one proposed here. 

\text{ }\\
\noindent\textbf{\normalsize{Discussion}}\\
In this paper, we examined the shape dependence of thermodynamic properties in quasi-equilibrium shape transformation processes. It is appropriate to use statistical methods as our system consists of many number of particles. We choose size of the longitudinal direction long enough to keep the number of particles high while particle density is low which allows Maxwell-Boltzmann statistics to be used. We showed that thermodynamic state functions strictly depend on the shape of the domain under strong confinement conditions.

The results are first presented numerically and then compared with the analytical results based on overlapped QBL method. For confined particles in between nested domains, rotation of the inner domain corresponds to the alteration of domain's shape. As a result, each rotational configuration gives distinct eigenvalue spectrum and leads to shape dependence in thermodynamic state functions. We suggest size-invariant shape transformation as a new thermodynamic process that can reduce both free energy and entropy of a system at the same time which is a rare situation (\textit{e.g.} seen in the formation of a snowflake) in nature\cite{seclaw}. This spontaneous decrease in entropy and free energy leads to unusual thermodynamic behaviors. For example, heat and work flows in the same direction during an isothermal shape transformation process and work output takes place at low temperature processes, different from the processes in conventional thermodynamic cycles.

The overlapped QBL method, proposed in this article, goes beyond the first two terms of PSF and Weyl conjecture and provides a way to predict the highly complicated third term in PSF. By use QBL methodology, the effective volume concept is introduced to take QSE and QShE into account explicitly. Once effective volume is properly determined in a domain, it's possible to predict the true behavior of thermodynamic properties under QSE and QShE. By means of overlapped QBL method we are able to predict and calculate shape dependence of thermodynamic properties without even solving Schr\"odinger equation or invoking any other mathematical tool other than simple geometric calculations. Quantum nature of the particles is embedded on the QBL thickness in this model. Combining this with the domain's geometric information, we obtain thermodynamic properties under QSE and QShE easily. Thus, the precise calculation of thermodynamic quantities is reduced into a simple geometric problem.

Note that these thermodynamic shape dependencies are not restricted to the domains considered here. Provided that the confinement domain consists of a space between nested objects and the confinement is strong so that QBLs overlap, similar thermodynamic behaviors are expected for any domain shape.

When inner square wire starts to rotate, one may need time-dependent solutions as the eigenvalue spectrum reorganizes itself in a finite-time with respect to boundary displacements \cite{qpmoving}. As a future work, we are planning to study a rotary quantum heat engine driven by the effect that is proposed here. Besides the rotational motion, translational motion could also be generated by the same mechanism, which would imply the possibility of designing linear motors in addition to rotary ones. QShE in Fermi and Bose gases are also under consideration. Taking particle-particle and particle-boundary interactions into account can be another extention of this problem. QShE on state functions may be considered as a macroscopic quantum phenomenon, since the effects persist even if one of the sizes is in macroscale. The results reported here may have a potential to open up new research topics and discussions in the newly growing quantum and nano thermodynamics field.

\text{ }\\
\noindent\textbf{\normalsize{Methods}}\\ 
\noindent\textbf{Numerical calculations.}
To take the confinement effects into account on thermodynamic state functions, we need to calculate them properly by considering the discrete nature of energy levels. Since there are no analytical solutions of Schr\"odinger equation for the confinement domains considered in this article, we obtain discrete sets of energy eigenvalues by solving Schr\"odinger equation numerically using finite element methods of \textsc{Comsol Multiphysics} and \textsc{Wolfram Mathematica} softwares. The boundaries are assumed to be impenetrable so that effects of confinement is maximized. Therefore, all boundary conditions are taken as Dirichlet. Each angular configuration of inner square leads to unique confinement domain shape. Schr\"odinger equation is solved and discrete sets of energy eigenvalues are obtained for each angular configuration (with $1^{\circ}$ steps). Then, Helmholtz free energy, entropy and internal energy of the system are calculated for each degree configuration.

In the calculation of torque, inner square is rotationally perturbed by a small amount of angle ($\Delta\theta=0.1$) and then, we divide free energy difference for this angular perturbation to the perturbation step, Eq. (6). To calculate total pressure force on the walls of inner square, small perturbations ($\Delta h=\SI{0.01}{\nano\metre}$) on the size of the square are done, and free energy responses to these perturbations are calculated. The perturbation steps are chosen as very small so that the results won't depend on the finiteness of the steps.

We made the error analysis of numerical results through the partition function from which we calculate all other thermodynamic quantities. Since there're infinitely many eigenvalues to be summed over, we truncate after a finite number of eigenvalues. For all calculations, we make sure that both truncation and mesh errors are much less than the contribution of the QShE corrections to the partition function.

In this article, solid curves in Figs. 2, 5, 7, 9 and 10 as well as density distributions in Fig. 3 are obtained using numerical calculations. During all calculations, particle density and temperature are taken as $5\times 10^{18} \text{cm}^{-3}$ and $300$K respectively. Longitudinal lengths of nested domains are chosen as $L_l\approx\SI{763}{\nano\metre}$ ($\alpha_l=0.05$). Outer and inner square sizes of nested square domain are $L_o\approx\SI{21}{\nano\metre}$ and $L_i\approx\SI{13}{\nano\metre}$ respectively ($\alpha_t=1$). For nested triangle and rectangle domains, sizes of outer and inner objects are determined by keeping confinement parameter of transverse direction unity.

\text{ }\\
\noindent\textbf{Analytical calculations.}
All thermodynamic state functions are represented by infinite summations over energy eigenvalues that are calculated from the Schr\"odinger equation. QSE corrections on thermodynamic properties can be obtained by several methods. One of them is using PSF which requires energy eigenvalues from the solution of Sch\"odinger equation for a particular confinement domain. A better methodology is Weyl conjecture that is based on precise enumeration of number of states for stationary Schr\"odinger equation and provides the same QSE terms for a domain with arbitrary shape, though it needs many integral calculations and does not give a physical explanation of QSE. On the other hand, usual QBL method not only generates the same QSE terms directly and simply, but also gives an explanation for underlying mechanism by providing a physical insight. Moreover, with the consideration of overlap regions, overlapped QBL method gives chance to analytically predict even QShE corrections.

The procedure of QBL method is quite straightforward to apply. Inside the partition function, instead of geometric volume, area or length (depending on the spatial dimension of the domain), one just need to write their effective values. One-dimensional (1D) illustration of different QBL approaches is shown in Fig. 10. QBL approaches are basically approximations to the density distribution of particles inside a confined domain. $0$th order approach approximates the density distribution by a homogenized region and two empty regions near to boundaries, whereas $1$st order approach approximates it by a narrower homogenized region and two linearly decreasing inhomogeneous regions near to boundaries. Although QBL approaches are constructed on a 1D system, they can directly be extended to higher dimensions.

\begin{figure}[t]
\centering
\includegraphics[width=0.49\textwidth]{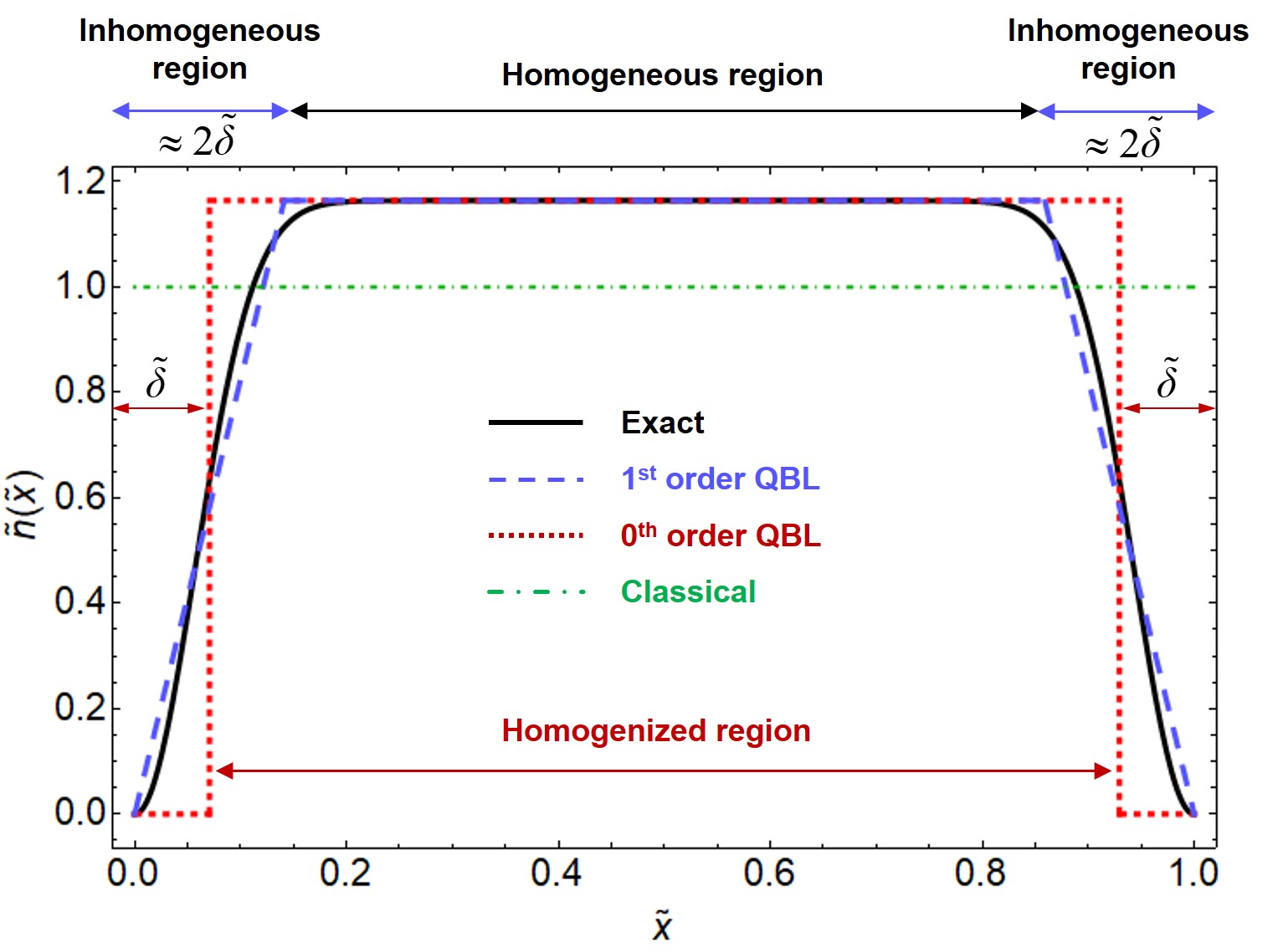}
\caption{\bfseries{Quantum boundary layer approaches for density distributions.} \mdseries Dimensionless density distribution of particles confined in a 1D domain. Solid black and dot-dashed green curves represent the exact and classical density distributions respectively. $1$\textsuperscript{st} order QBL denoted by dashed blue lines is the linear approximation to the inhomogeneous region with thickness $2\tilde{\delta}$ in exact density distribution. $0$\textsuperscript{th} order QBL is represented by dotted red lines where the exact density distribution is approximated by empty regions with thickness $\tilde{\delta}$ near to boundaries and homogenized density region in the remaining parts. Here $\tilde{\delta}=\delta/L$ and $\tilde{x}=x/L$ where $x$ is the position and $L$ is the length of the 1D domain.}
\label{fig:pic10}
\end{figure}

When confinement is too strong and domain shape is appropriate so that QBLs of different boundaries overlap on each other, then those regions are subtracted twice in the usual calculation of effective sizes. Therefore, the overlap regions has to be added for a proper calculation of effective sizes. In overlapped QBL method proposed in this article, 0th order QBL approach is used (dotted red curve in Fig. 10).

Now let's focus on the analytical calculation of the summation in the partition function (Eq. 1) for the nested square confinement domain (Fig. 1). Eigenvalues for that summation are normally found numerically from the solution of Schr\"odinger equation. For the analytical calculation of the partition function, we replace the summation term with the conventional integral form of it and replace the geometric volume term ($V=(L_o^2-L_i^2)L_l$) with the effective volume term as follows:
\begin{equation}
\begin{split}
\left(\frac{\sqrt{\pi}}{2\alpha_l}-\frac{1}{2}\right)\sum_k \exp\left(-\frac{\alpha_t^2}{\pi^2}\tilde{k}^2\right)\approx\frac{\pi^{3/2} V_{\mli{eff}}}{8L_c^3},
\end{split}
\end{equation}
where the effective volume is $V_{\mli{eff}}=[(L_o-2\delta)^2-(L_i+2\delta)^2](L_l-2\delta)+V_{\mli{ovr}}=(L_{o,\mli{eff}}^2-L_{i,\mli{eff}}^2)L_{l,\mli{eff}}+V_{\mli{ovr}}$ and here, overlap volume can analytically be obtained as a function of $\theta$ by using simple geometric relations as follows,
\begin{equation}
\begin{split}
V_{ovr}=
\begin{cases}
	0,   & \text{for} \;  0^{\circ}\leq\theta<\theta_{*} \\
	L_{l,\mli{eff}}\frac{\tan\theta}{2}\left[L_{i,\mli{eff}}\left(1+\frac{1}{\tan\theta}\right)-\frac{L_{o,\mli{eff}}}{\sin\theta}\right]^2,   & \text{for} \;  \theta_{*}\leq\theta\geq45^{\circ},
\end{cases}
\end{split}
\end{equation}
where $\theta_{*}=\text{Round}\left[\frac{180}{\pi}\arctan\left(\frac{L_{o,\mli{eff}}\sqrt{2L_{i,\mli{eff}}^2-L_{o,\mli{eff}}^2}-L_{i,\mli{eff}}^2}{L_{i,\mli{eff}}^2-L_{o,\mli{eff}}^2}\right)\right]$. 

Therefore, temperature sensitivity of overlap volume, which is needed for the analytical calculation of entropy and internal energy (Eq. 4 and 5), can be obtained as
\begin{equation}
\begin{split}
\frac{\partial V_{\mli{ovr}}}{\partial T}=-\sqrt{2\tan\theta}\left(1+\cot\frac{\theta}{2}\right)L_{l,\mli{eff}}\frac{\delta}{T}\sqrt{\frac{V_{\mli{ovr}}}{L_{l,\mli{eff}}}}+\frac{\delta}{T}\frac{V_{\mli{ovr}}}{L_{l,\mli{eff}}}.
\end{split}
\end{equation}
The first term is the dominant term as long as the longitudinal size of the domain is much larger than $L_{*}$, which is in fact the necessary condition to use Maxwell-Boltzmann statistics.

The torque can now analytically be calculated by taking the $\theta$ derivative of Eq. (3) in which all terms become analytical by use of Eqs. (2) and (10). In a similar manner, analytical expression for the pressure forces exerted on the inner walls can be obtained by $Nk_BT(L_{w,\mli{eff}}L_{l,\mli{eff}}/V_{\mli{eff}})=4Nk_BT/[L_{o,\mli{eff}}(\sin\theta+\cos\theta)-L_{i,\mli{eff}}]$, where $L_{w,\mli{eff}}$ represents the effective transverse length of the inner square that pressure force exerts. Ratio of this torque to pressure force gives the application points of forces analytically. Fig. 6 and dotted/dashed curves in Figs. 2, 5, 7a, 7b and 7c in this article are obtained using analytical calculations.

\flushleft{Date of Submission: \today}\\
\text{ }\\
\noindent  \textbf{Acknowledgments}\\
\begin{scriptsize}
\noindent We acknowledge kind support from Istanbul Technical University.
\end{scriptsize}
\par
\text{ }\\
\noindent  \textbf{Author contributions}\\
\begin{scriptsize}
\noindent Both authors contributed to all aspects of this work.
\end{scriptsize}
\par
\text{ }\\
\noindent  \textbf{Additional information}\\
\begin{scriptsize}
\noindent Correspondence and requests for materials should be addressed to A. Sisman.
\end{scriptsize}
\par
\text{ }\\
\noindent  \textbf{Competing financial interests}\\
\begin{scriptsize}
\noindent The authors declare no competing financial interests.
\end{scriptsize}
\end{document}